\documentclass{iau_FM}
\usepackage{graphicx}

\title[FM4 - Planetary Nebulae as tracers in nearby clusters] 
{Extended halos and intracluster light using \\ Planetary Nebulae as
  tracers in nearby clusters}

\author[Magda Arnaboldi]   
{Magda Arnaboldi$^{1,2}$}

\affiliation{$^1$ European Southern Observatory,
  K. Schwarszchild-str. 2, 85748 Garching, Germany \\ email: {\tt
    marnabol@eso.org} \\[\affilskip] $^2$INAF, Oss. Astr. di Pino
  Torinese, 10025 Pino Torinese, Italy}

\pubyear{2015}
\jname{Extragalactic Planetary Nebulae} 
\editors{L. Stanghellini, ed.}
\begin{document}

\maketitle

\begin{abstract}
Since the first detection of intracluster planetary nebulae in 1996,
imaging and spectroscopic surveys identified such stars to trace the
radial extent and the kinematics of diffuse light in clusters. This
topic of research is tightly linked with the studies of galaxy
formation and evolution in dense environment, as the spatial
distribution and kinematics of planetary nebulae in the outermost
regions of galaxies and in the cluster cores is relevant for setting
constraints on cosmological simulations. In this sense, extragalactic
planetary nebulae play a very important role in the near-field
cosmology, in order to measure the integrated mass as function of
radius and the orbital distribution of stars in structures placed in
the densest regions of the nearby universe.

\end{abstract}

\firstsection 

\section{Introduction}\label{sec0}
Hierarchical structure formation theories predict that massive galaxy
clusters are built through the infall of matter, i.e.  galaxies,
groups and sub-clusters, along large scale filaments (\cite[White \&
  Rees (1978)]{WR+78}). Since mass accretion is still active today,
nearby galaxy clusters may be at different epochs of their mass
assembly, and they may still be in an unmixed state.

The gravitational forces acting on galaxies during cluster formation
and its evolution unbind a fraction of their stars, which then end-up
orbiting in the intracluster region. This cluster-driven diffuse
stellar component is often referred to as ``{\it intracluster light
  (ICL)}''.  Numerical simulations show that the amount of ICL in
clusters depends on the cluster mass and dynamical state (\cite[Cooper
  \etal\ 2015]{Coo+15}). A more massive, older, dynamically evolved
cluster may contain a larger amount of ICL than a less massive or a
dynamically younger system (\cite[Rudick \etal\ 2006]{R+06};
\cite[Murante \etal\ 2007]{M+07}). Furthermore, in a highly evolved
dynamically old cluster, the ICL morphology would be more diffuse, with
relative few streams, while a dynamically young cluster dominated by
groups still in the process of merging is likely to be dominated by
ICL found in streams (\cite[Rudick \etal\ 2009]{R+09}). Therefore the
study of the amount, distribution and kinematics of the ICL may
provide information on the cluster accretion history and evolutionary
state, as well as about the evolution of cluster galaxies.

Several studies have mapped the distribution of the ICL in nearby
clusters (see \cite[Mihos 2015]{Mihos15} for a review): since the
serendipitous discovery of three intracluster planetary nebulae
(ICPNs) moving at $v_{mean} = 1400$ kms$^{-1}$ along the line-of-sight
of NGC 4406 ($v_{sys} = -240$kms$^{-1}$) in the Virgo cluster core
(\cite[Arnaboldi \etal\ 1996]{arna+96}), these stellar tracers have been
used to map the spatial distribution and kinematics of the diffuse
light in clusters. In what follows, the physical properties of the
ICPNs samples in the nearby clusters, i.e. Virgo, Hydra~I and Coma,
will be outlined and compared.

\section{The Virgo cluster}\label{sec1}
Several studies investigated the properties of the ICL in the core of
the nearby Virgo cluster (Feldmeier \etal\ 1998, 2003, 2004; Arnaboldi
\etal\ 2002, 2003; Aguerri \etal\ 2005) and around M49 (\cite[Feldmeier
  \etal\ 2004]{F+04}). Expanding on these earlier survey works,
\cite[Castro-Rodriguez \etal\ (2009)]{Castro+09} completed a survey
campaign of the ICL distribution on larger scales, outside the center
of the Virgo cluster. In total, they covered more than 3 square
degrees in Virgo, at eleven different positions in the cluster and at
distances between 80 arcmin and some 100 arcmin from the Virgo cluster
center. In several of these fields, the ICL is at least two magnitudes
fainter than in the Virgo core region.

These new results are in agreement with observations of intermediate
redshift clusters (\cite[Zibetti \etal\ 2005]{Zibetti+05};
\cite[D'Souza \etal\ 2014]{DS+2014}), and with the results of high
resolution hydrodynamical simulations of cluster formation in a
$\lambda$ CDM universe (see also \cite[Murante \etal\ 2004]{M+04}),
which predict that the ICL is more centrally concentrated than cluster
galaxies and that the largest portion of the ICL is formed during the
assembly of the most luminous cluster galaxies.

\section{Kinematics of planetary nebulae in the Virgo cluster core}\label{sec2}
Early attempts to measure spectra for ICPN in Virgo (Arnaboldi
\etal\ 2003, 2004; Doherty \etal\ 2009) provided spectra for 52 single
PNs around M87 and the Virgo core. These single spectra showed the
[OIII]$\lambda$4959/5007\AA\ double emissions, confirming the
identification of these emission line candidates and their association
with the Virgo cluster. Furthermore the measured expansion velocities
of the [OIII] envelope for these bright PNs is similar to those
measured for the bright PNs in the Milky Way bulge (\cite[Arnaboldi
  \etal\ 2008]{arna+08}).

The sparsely sampled projected phase-space of these PNs that was built
from their line-of-sight velocity ($v_{LOS}$) vs. radial distance from
the center of M87 was probing regions with different kinematics. The
halo of M87 was clearly detected as a high density region centered
around its systemic velocity (1300 kms$^{-1}$), and a broad
line-of-sight velocity distribution (LOSVD) was measured at distances
of 2000 arcsec and larger. Such broad LOSVD is reminiscent of the
velocity distribution of the galaxies in the Virgo cluster
(\cite[Binggeli \etal\ 1987]{bing+87}).

The recent, extended PN survey carried out by \cite[Longobardi
  \etal\ (2013)]{Long+13} represents a major increase with respect
with previous sample in the same M87 area of a factor 15. With the
narrow band imaging carried out with Suprime-Cam at Subaru, they
detected 688 PN candidates in an area of nearly 0.5 sq-deg covering
M87. The spectroscopic follow-up was performed with Flames at UT2 on
the VLT; 287 spectra were obtained for single PNs (\cite[Longobardi
  \etal\ 2015]{Long+15}). This sample was analyzed co-jointly with 12
PNs from \cite[Doherty \etal\ (2009)]{D+09} in the same area, giving a
merged total sample of 301 PNs.

\cite[Longobardi \etal\ (2015)]{Long+15} built the LOSVD for the
entire sample of the spectroscopically confirmed PNs. The LOSVD displays
broad asymmetric wings that cannot be matched by a single Gaussian,
see Figure~5 in \cite[Longobardi \etal\ (2015)]{Long+15}.  The entire
LOSVD was fitted by a narrow Gaussian centered at the systemic velocity
of M87 and $\sigma_n \simeq 300$ kms$^{-1}$ and a broad Gaussian
component with average velocity at $v_{b}\simeq 1000 $ kms$^{-1}$ and
$\sigma_{b} \simeq 900$ kms$^{-1}$. This kinematical decomposition is
carried out in the projected space space, $v_{LOS}$ vs. $R_{Major}$,
and PNs can be assigned either to the narrow Gaussian, i.e. the M87
halo component, or to the broad component, i.e. the ICL, via a robust
sigma-clipping procedure (\cite[Longobardi \etal\ 2015]{Long+15}). 

Such kinematically identified PN populations showed different physical
properties, in addition. Firstly, the M87 halo PNs number density
profile follows the Sersic fit to the V-band surface brightness
profile from \cite[Kormendy \etal\ (2009)]{K+09}, while the ICPNs
number density profile is consistent with a power law, $N_{ICPN}(R)
\propto R^\gamma$ with $\gamma = [-0.34, -0.04]$. Secondly, the two PN
populations have different specific frequencies with the ICL having
three times larger PN yield per bolometric luminosity than the M87
halo light. Thirdly, the morphology of the PN luminosity function
(PNLF) for these two components differs. The PNLF for the M87 halo
sample has a steeper gradient than the M31 PNLF or the standard
Ciardullo's analytical formula (\cite[Ciardullo
  \etal\ 1989]{Ciardullo+89}), while the PNLF for the ICPN shows a
``dip'' which is reminiscent of a similar morphological feature
detected in the SMC/LMC PN samples (\cite[Jacoby \& De Marco
  2002]{SMC02}, \cite[Reid \& Parker 2010]{LMC2010}). The similarity
of the IC PNLF with those observed for low-luminosity, low
metallicity, star forming irregulars support the identification of
these galaxies as likely progenitors of the stars that are now found
associated with the ICL in the Virgo cluster core (\cite[Longobardi
  \etal\ 2015]{Long+15}).

\section{Detecting Planetary Nebulae beyond 20 Mpc distance}\label{sec3}
The brightest PNs at 50 Mpc distance have fluxes of
$7.8\times10^{-18}$ erg~s$^{-1}$cm$^{-2}$ ($\sim 7$ photons/min on 8m
tel.). In the Coma cluster, the fluxes of the brightest PNs are four
times fainter. Hence the PN [OIII]$\lambda$5007\AA\ emission cannot be
detected using a narrow band filter because the sky noise in a $30-40$
\AA\ window centered on the red-shifted [OIII] PN emission is of the
same order as the signal from such line. A step forward for the
detection of PN in elliptical galaxies in clusters at distances larger
than 20 Mpc is to decrease significantly the sky noise. This is
achieved with the Multi Slit Imaging Spectroscopy Technique (MSIS;
\cite[Gerhard \etal\ 2005]{ger+05}).

MSIS is a blind search technique, that combines a mask of parallel
multiple slits with a narrow band filter, centered on the red-shifted
[OIII]$\lambda$5007\AA\ line at the Hydra~I/Coma mean systemic
velocities, to obtain spectra of all PNs that lie behind the slits
(\cite[Gerhard \etal\ 2005]{ger+05}).  The sky noise at the PN [OIII]
emission line now comes from a spectral range of only a few \AA,
depending on slit width and spectral resolution (\cite[Arnaboldi
  \etal\ 2007]{arna+07}). Several tens of PNs were detected using the
MSIS observations in the Hydra~I (\cite[Ventimiglia
  \etal\ 2011]{ven+11}) and Coma (\cite[Gerhard \etal\ 2007]{ger+07},
\cite[Arnaboldi \etal\ 2007]{arna+07}) clusters.

\begin{figure}
\center\includegraphics[width=10.0cm]{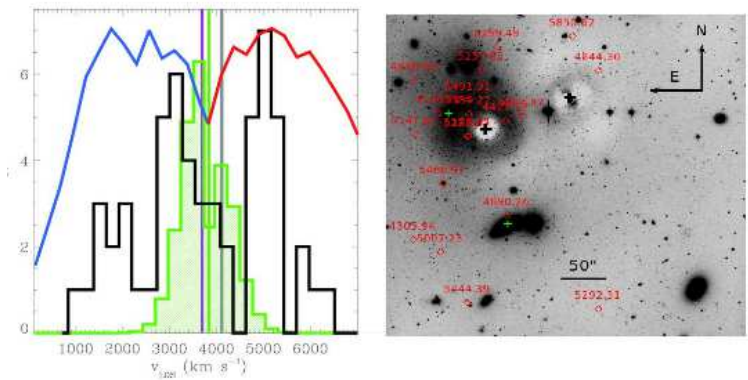}
\caption{Left panel: the PNs LOSVD in the core of the Hydra~I cluster
  obtained from the MSIS observations by \cite[Ventimiglia
    \etal\ (2011)]{ven+11}. The LOSVD is shown by the black line,
  while the blue and red curves indicate the measured transmission
  curves of the narrow band filters used to cover the Hydra
  velocities; the normalization is arbitrary. The magenta, green and
  gray vertical lines indicate the systemic velocity of the Hydra-I
  cluster, NGC~3311 and NGC~3309, respectively. The green histogram
  shows the simulated MSIS LOSVD for a Gaussian velocity distribution
  with $\sigma_{core} = 464$ kms$^{-1}$; from Ventimiglia et
  al. (2011). Right panel: Positions and velocities of PNs (red
  diamonds); labels indicate the PNs’ $V_{LOS}$ in kms$^{-1}$
  associated with the red peak at 5000 kms$^{-1}$ in the PN LOSVD of
  the Hydra~I cluster. The red peak PNs are superposed on the
  residual V-band image, which shows the substructures in the diffuse
  light in the Hydra I cluster core. The black crosses indicate the
  position of NGC 3311 (center) and NGC 3309 (upper right),
  respectively. The green crosses indicate HCC~026 and HCC~007. The
  FoV is $6.8'\times 6.4'$. From \cite[Arnaboldi
    \etal\ (2012)]{arna+12}. }
\label{figure1}
\end{figure}

{\it The Hydra~I cluster and NGC~3311} - \cite[Ventimiglia
  \etal\ (2011)]{ven+11} performed the MSIS observations with FORS2 at
the ESO VLT of the core region of the Hydra~I cluster, centered on its
cD galaxy, NGC~3311.  They detected a total of 56 PNs in a single
field of $100 \times 100$ kpc$^2$ , and analyzed the LOSVD.  The PNs
LOSVD in this region has several peaks, see Figure~\ref{figure1}: in
addition to a broad symmetric component centered at the systemic
velocity of the cluster ($v_{Hydra} = 3683$ kms$^{-1}$), two narrow
peaks are detected at $1800$ kms$^{-1}$ and $5000$ kms$^{-1}$. These
secondary peaks unmask the presence of un-mixed stellar populations in
the Hydra~I core.

The spatial distribution of the PNs associated with the narrow
velocity sub-component at $5000$ kms$^{-1}$ is superposed and
concentrated on an excess of light in the North-East quadrant of
NGC~3311, as detected from the two-dimensional light decomposition of
the NGC~3311/NGC~3309 B band photometry, see \cite[Arnaboldi
  \etal\ (2012)]{arna+12}. On the same sky location, we detect a group
of dwarf galaxies with $v_{LOS} \approx 5000$ kms$^{-1}$.  Deep
long-slit spectra obtained at the position of the dwarf galaxy HCC~026
situated at the center of the light excess show absorption line
features from both HCC~026 and the light excess which are consistent
with $v_{LOS} \sim 5000$ kms$^{-1}$ . \cite[Arnaboldi
  \etal\ 2012]{arna+12} concluded that the PNs in the $5000$ kms$^{-1}$
sub-component, the dwarfs galaxies and the light excess in the
North-East quadrant of NGC~3311 occupy the same region of the
phase-space, and are physically associated.

Hence also in the case of the relaxed Hydra-I cluster, PN kinematics,
photometry and deep absorption spectra support the evidence for an
accretion event whose stars are being added to both the cluster core
and the halo of NGC~3311.

\section{The on-going sub-cluster merger in the Coma cluster core}
\begin{figure}
\center\includegraphics[width=8.0cm]{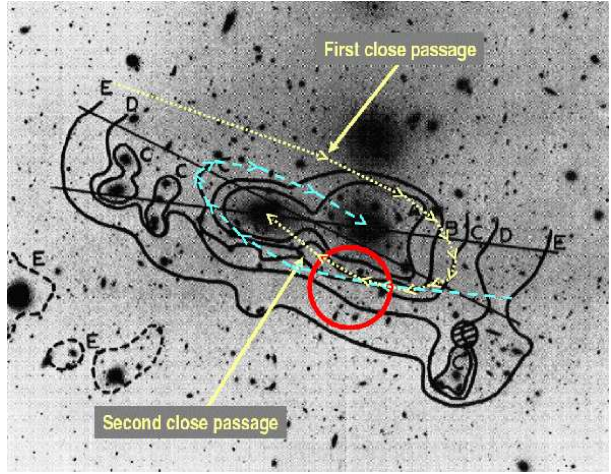}
\caption{The position of the MSIS field (red circle) studied in
  \cite[Gerhard \etal\ (2007)]{ger+07} on the diffuse light isodensity
  contours drawn by \cite[Thuan \& Kormendy 1977]{TK+77} in the Coma
  cluster core. The MSIS field is about $5'$ south of NGC 4874. The
  second Coma cD galaxy NGC 4889 is $7'$ east (to the left) of NGC
  4874. The bright object north of NGC~4874 is the star which
  prevented \cite[Thuan \& Kormendy 1977]{TK+77} from reliably
  determining the northern parts of the isodensity curves in their
  photographic photometry. Note the strong elongation of the
  distribution of ICL in the Coma cluster core. The likely orbits of
  NGC~4889 and NGC~4874 up to their present positions are sketched by
  the yellow dotted and magenta dashed lines, respectively; see
  \cite[Gerhard \etal\ (2007)]{ger+07}. }
\label{figure2}
\end{figure}

The Coma cluster is the richest and most compact of the nearby
clusters, yet there is growing evidence that its formation is still
on-going.  A sensitive probe of this evolution is the dynamics of
intracluster stars, which are unbound from galaxies while the cluster
forms, according to cosmological simulations.  With the MSIS
technique, \cite[Gerhard \etal\ (2005)]{ger+05} detected and measured
the $v_{LOS}$ of 37 ICPNs associated with the diffuse stellar
population of stars in the Coma cluster core, at 100 Mpc
distance. These are the most distant single stars whose spectra were
acquired, in addition to cosmological supernovae stars.  \cite[Gerhard
  \etal\ (2007)]{ger+07} detected clear velocity sub-structures within
a 6 arcmin diameter field centered at $\alpha(J2000) = 12:59:41.8;\,
\delta(J2000) 27:53:25.4$, nearly coincident with the field observed
by \cite[Bernstein \etal\ (1995)]{Bern+95} and $\sim 5$ arcmin away
from the cD galaxy NGC~4874.  A sub-structure is present at $\sim$
5000 ${\rm ~kms^{-1}}$, probably from in-fall of a galaxy group, while
the main intracluster stellar component is centered around $\sim$6500
${\rm ~kms^{-1}}$, $\sim$700 ${\rm ~kms^{-1}}$ offset from the nearby
cD galaxy NGC~4874 ($v_{sys} = 7224$ kms$^{-1}$; from NED) . The
kinematics and the elongated morphology of the intracluster stars
(\cite[Thuan \& Kormendy 1977]{TK+77}) show that the cluster core is
in a highly dynamically evolving state. In combination with galaxy
redshift and X-ray data, this argues strongly that the cluster is
currently in the midst of a sub-cluster merger.  The NGC~4889
sub-cluster is likely to have fallen into Coma from the eastern A2199
filament, in a direction nearly in the plane of the sky, meeting the
NGC~4874 sub-cluster arriving from the west. The two inner sub-cluster
cores are presently beyond their first and second close passage,
during which the elongated distribution of diffuse light has been
created, see Figure~\ref{figure2}.  \cite[Gerhard
  \etal\ (2007)]{ger+07} also predicted the kinematic signature expected
in this scenario, and argued that the extended western X-ray arc
recently discovered traces the arc shock generated by the collision
between the two sub-cluster gaseous halos.

\section{Summary and Conclusions}
The kinematics of the diffuse light mapped via by extragalactic PNs
provide unique information to asses the dynamical status of the nearby
cluster cores. In all cases the kinematical data indicate that the
galaxy halos and ICL are discrete components and the former do not
blend continuously in the latter. The evidence for merging (in Coma)
and accretion on the extended halos (M87, NGC 3311) indicates that the
build up of the diffuse light in the cluster cores is an on-going
process.

The predicted spatially averaged radial distribution of ICL from
recent high resolution hydrodynamical simulations of cluster formation
in $\Lambda$ CDM universe is in broad agreement with the observed
radial profiles for the ICL in clusters.  Furthermore predictions from
these simulations indicate that the more massive progenitors dominate
the diffuse light in simulated cluster close to the center
(\cite[Murante \etal\ 2007]{M+07}, \cite[Puchwein \etal\ 2010]{P+10}).
The prediction that the largest portion of the ICL is formed during
the assembly of the most luminous cluster galaxies is supported by the
observed ongoing mergers in Virgo, Hydra~I and Coma cores.

Quantitative analysis of the ICL kinematic from cosmological
simulations is on-going. Studies of the galaxy halo and ICL particles
in cosmological simulations (\cite[Dolag \etal\ 2010]{Dolag+10},
\cite[Cui \etal\ 2014]{Cui+14}) further support the physical
distinction between the central bright galaxy and the ICL component in
clusters.

\section{Acknowledgment}
The author would like to thank the organizers for the invitation to
give a review on extragalactic PNs as tracers of diffuse light at the IAU
FM 4.

\end{document}